\documentclass[%
superscriptaddress,
 aip,
 jcp,
 amsmath,amssymb,
 reprint,%
citeautoscript
]{revtex4-2}

\usepackage{subfigure}
\usepackage{hyperref}
\usepackage{graphicx}
\usepackage{dcolumn}
\usepackage{bm}
\usepackage[utf8]{inputenc}
\usepackage[T1]{fontenc}
\usepackage{mathptmx}
\usepackage{tikz}
\usetikzlibrary{positioning,calc}
\usetikzlibrary{trees}
\usetikzlibrary{decorations.pathmorphing}
\usetikzlibrary{decorations.markings}
\tikzset{fontscale/.style = {font=\relsize{#1}}    }
\usepackage[version=3]{mhchem}
\usepackage{color}
\definecolor{mma1}{rgb}{0.3725,0.5098,0.7020}
\definecolor{mma2}{rgb}{0.8745,0.6078,0.2039}
\definecolor{mma3}{rgb}{0.507813,0.714844,0.2039}
\definecolor{mma4}{rgb}{0.9137,0.3882,0.2398}

\begin{document}

\title{Stochastic scattering theory for excitation induced dephasing: Comparison to the Anderson-Kubo lineshape}
\author{Hao~Li}
\affiliation{Department of Chemistry, University of Houston, Houston, Texas 77204, United~States}

\author{Ajay~Ram~Srimath~Kandada}
\affiliation{Department of Physics and Center for Functional Materials, Wake Forest University, 
1834 Wake Forest Road, Winston-Salem, North Carolina~27109, United~States}

\author{Carlos~Silva}
\affiliation{School of Chemistry and Biochemistry, Georgia Institute of Technology, 901 Atlantic Drive, Atlanta, GA~30332, United~States}
\affiliation{School of Physics, Georgia Institute of Technology, 837 State Street, Atlanta, GA~30332, United~States}
\affiliation{School of Materials Science and Engineering, Georgia Institute of Technology, North Avenue, Atlanta, GA~30332, United~States}

\author{Eric~R.~Bittner}
\email{ebittner@central.uh.edu}
\affiliation{Department of Chemistry, University of Houston, Houston, Texas 77204, United~States}

\date{\today}

\begin{abstract}
In this paper we present a quantum stochastic model for 
spectroscopic line-shapes in the presence of a co-evolving and non-stationary background population of excitations. Starting from a field theory
description for interacting bosonic excitons, we derive a reduced model 
whereby optical excitons are coupled to an incoherent background
via scattering as mediated by their screened Coulomb coupling.
The Heisenberg equations of motion for the optical excitons 
are then driven by an auxiliary stochastic population variable, which 
we take to be the solution of an Ornstein-Uhlenbeck process.  It\^o's Lemma
then allows us to easily construct and evaluate correlation functions
and response functions.   Focusing on the linear response, we compare our
model to the classic Anderson-Kubo model.  While similar in motivation, 
there are profound differences in the predicted lineshapes, notably in terms
of asymmetry, and variation with increasing background population.
\end{abstract}

\maketitle

\section{Introduction}

One of the cornerstones of modern
spectroscopy is that the lineshape of a
spectral transition gives an indication 
of the underlying environment and background
dynamics of the system being probed. 
According to the Anderson-Kubo model (AK), \cite{Anderson:JPSJ1954,Kubo:JPSJ1954} the
energy levels of a molecule or atom are modulated
by fluctuations within the surrounding
environment.  Such fluctuations can arise
from nuclear and electronic motions of the 
surrounding environment which induce a noisy
driving field.   A suitable model for this is 
to write that a transition frequency
has an intrinsic time dependence 
\begin{align}
    \omega(t) = \omega_0 + \delta \omega(t)
\end{align}
where $\omega_0$ is the central (mean) transition frequency and $\delta \omega(t)$
is some time-dependent modulation with $\langle\delta\omega(t)\rangle = 0$.
Lacking detailed knowledge of the environment,
it is reasonable to write the frequency auto-correlation 
function in terms of the deviation about the 
mean, $\Delta$ and a correlation time, $ \tau_c = \gamma^{-1}$, {\em viz.}
\begin{align}
    \langle \delta\omega(t)\delta\omega(0)
    \rangle = \Delta^2 e^{-\gamma   t}.
\end{align}
From this we can go on to write the 
linear response for light absorption or emission between initial and final 
quantum states.
The model has two important limits. 
First, if $\Delta\gamma \ll 1$ 
absorption line shape
takes a Lorenzian functional
form with a homogeneous width determined by the dephasing time 
$T_2^{-1} = \Delta^2/\gamma$. 
On the other hand, if $\Delta \gamma \gg 1$, 
the absorption spectrum takes a Gaussian form with a line width 
independent of the correlation time.  In this limit, fluctuations
are slow and the system samples a broad distribution of environmental motions. 
Increasing the rate of the fluctuations (i.e. decreasing the 
correlation time) leads to the effect of motional narrowing
where by the line width becomes increasingly narrow. 

All of this assumes that the background 
dynamics are more or less due to 
fluctuations about a stationary state. 
This is certainly the case for 
isolated chromophores embedded in 
a condensed phase environment. However
for a semiconducting system, 
one can have weak Coulomb interactions
between excitations as well as 
a nonstationary ensemble of background
excitations produced by
broad-band excitation from an initial 
laser pulse. 
Such transient  
fluctuations in the number density 
of excitations, $\delta n(r,t)$ 
induce space-charges and hence fluctuating potential.\cite{Lindberg1994,Perakis1996}  
This is effect is known as excitation induced dephasing (EID)  
and has been observed in a number of contexts.
\cite{Hu1994,Schneider2004,Dunbar2002,Kaindl2001,siemens2010resonance,Schultheis1986,Katsch2020}

In this paper, we lay the groundwork for an accompanying 
paper concerning 
the observation of the phenomena of 
excitation induced dephasing (EID) in the 
coherent 2D spectroscopy of a hybrid perovskite semiconductor
\ce{(PEA)2PbI4} (PEA = pheny\-ethyl\-ammonium) --- a multiple-quantum-well-like single-layer metal-halide perovskite derivative. We choose this material to test the theoretical framework developed 
here because of its susceptibility to strong many-body
effects~\cite{Kato2003,Thouin2018,thouin2019enhanced} and dynamic 
exciton-lattice coupling that drives their dynamics~\cite{Neutzner2018,thouin2019phonon,thouin2019polaron,SrimathKandada2020}
In this system, we reported that the homogeneous linewidth broadened 
with increasing pumping fluence, which is a hallmark of EID.  Moreover, 
we reported that this line width progressively narrows as the
exciton population evolves in time and we  attributed this to the 
transient decay of background exciton population. 
In this paper, we start with a generalized field theory for interacting bosonic excitons.
We then reduce this to a quantum stochastic model whereby the 
background population evolves from a non-stationary initial population 
generated by the initial excitation.   Here we focus upon the 
linear response and compare our approach to AK. In Ref.~\citenum{paper1}
we shall consider the non-linear responses and compare the predictions of our
approach to experimental coherent 2D spectroscopic signals.

\section{ Many-body model}
Here we consider the case where we have an ensemble of excitons which we will write first 
in terms of field operators $\hat\psi^\dagger(r)$ and $\hat\psi(r)$ which create and remove 
excitons at location $r$.  These are bosonic operators $[\hat\psi(r'),\hat\psi^\dagger(r)] = \delta(r-r')$.
\begin{align}
    H &= \int \frac{\hbar^2}{2m}(\nabla \hat\psi^\dagger)(\nabla\hat \psi) dr \\
    &+ \frac{1}{2}\int dr dr'
    \hat\psi^\dagger(r')\hat\psi^\dagger(r) V(r-r') \hat\psi(r')\hat\psi(r)
\end{align}
By Fourier transform we define
\begin{align}
    \hat\psi(r) = \frac{1}{\sqrt{V}}\sum_k a_k e^{ik.r}
\end{align}
and recast the Hamiltonian as 
\begin{align}
    H = \sum_k \frac{\hbar^2 k^2}{2m}a_k^\dagger a_k  + \frac{1}{2}\sum_{kk'q}V_q a^\dagger_{k+q}a^\dagger_{k'-q}a_ka_{k'}
\end{align}
where $V = L^3$ is the unit volume and 
\begin{align}
    V_q = \int V(r) e^{iq.r}dr
\end{align}
is the Fourier component of the many-body interaction potential.  In general, $V(r)$ always will 
include short and long-ranged contributions.

Here we have taken only the $\ell = 0$ (s-wave) term in the expansion. 
The interaction potential $V(r)$ always will 
include short and long-ranged contributions. 
However, for any finite-ranged potential,
we can express this interaction in terms of the 
$s$-wave scattering length, $a$, according to the Born approximation
$  V_o = {4\pi \hbar^2 a}/{\mu}$.
Such an approximation is valid in the limit that the momentum exchange is small 
compared to the effective range of the potential.
Under this approximation, we can replace the actual exciton/exciton interaction 
potential with an arbitrary, but smooth, fictitious potential that has the same value of $V_o$.
For the case at hand, we assume that the exciton/exciton interaction takes the form of a screened-Coulomb or Yukowa potential of the form ($r_c = 1/\alpha$ is the screening length)
\begin{align}
    V(r) =\frac{1}{4\pi}\frac{e^{-\alpha r}
    }{r}, 
\end{align}
the Fourier-transform of which reads
\begin{align}
    V(q) =\frac{1}{q^2 + \alpha^2}.
\end{align}
The full exciton $H$ is then
\begin{align}
    H = \sum_k \frac{\hbar^2 k^2}{2m}a_k^\dagger a_k  + \frac{1}{2}\sum_{kk'q}V_q a^\dagger_{k+q}a^\dagger_{k'-q}a_ka_{k'}.
\end{align}
We now split out the $k=0$ excitons and treat the $k\ne 0$ excitons as a bath. 
First, one finds that Hamiltonian can be re-written as
\begin{align}
H &=    \hbar\omega_0 a^\dagger_0a_0 + \frac{V_0}{2} a^\dagger_0 a^\dagger_0a_0a_0 \nonumber \\
  &+\frac{V_0}{2}\sum_{q\ne 0}(4 a^\dagger_0 a^\dagger_qa_0a_q + a^\dagger_q a^\dagger_{-q}a_0a_0 + a^\dagger_0 a^\dagger_0a_qa_{-q})
\end{align}
where $\hbar\omega_k = \hbar^2 k^2/2m$ is the exciton dispersion 
and the second term is the $k=0$ exciton self-interaction. 
The first $q\ne 0$ interaction term arises from direct and 
exchange interactions between the $k=0$ excitons and the $k\ne 0$
excitons.  The next two terms correspond to exciton pair creation/annihilation.  
Within the Bogoliubov theory
of an interacting Bose condensate, these terms give rise to a 
linearization of the energy dispersion around $k=0$.
Whereas in the Bogoliubov approach, one can assume that the $k=0$ population is very large and use this to 
re-write the interaction in terms of the $k=0$ populations by 
writing $\langle a_0 a_0\rangle = \langle a_0^\dagger a_0^\dagger\rangle \approx n_0$ and take $n_o$ to approach macroscopic populations, in the case at hand we need to keep
these as quantum operators and we will treat the $q\ne 0$ operators as sources of quantum noise and dissipation.

Rewriting the Hamiltonian once more, we collect all the $k\ne 0$ terms 
\begin{align}
    H &=   \hbar\omega_0 a^\dagger_0a_0 + \sum_{k\ne 0}\hbar\omega_q a_q^\dagger a_q + \frac{V_0}{2} a^\dagger_0 a^\dagger_0a_0a_0  \nonumber \\
      & +a^\dagger_0 a_0 \left[{2V_0}\sum_{q\ne 0}(a^\dagger_qa_q)\right] \nonumber \\
      &+a^\dagger_0 a^\dagger_0\left[ \frac{V_0}{2}\sum_{q\ne 0} a_q a_{-q}\right]
       +a_0 a_0\left[ \frac{V_0}{2}\sum_{q\ne 0} a_q^\dagger a_{-q}^\dagger \right] 
       \label{eq:hfinal}
\end{align}
We can use the form of this Hamiltonian 
to derive quantum stochastic equations of motion for the $k=0$ operators treating the $k\ne 0$ terms as a 
Markov bath. 
First, define the $k=0$ term as 
\begin{align}
    H_0 &=  \hbar\omega_0 a^\dagger_0a_0 +  \frac{V_0}{2} a^\dagger_0 a^\dagger_0a_0a_0 
    \nonumber \\
    &+2 V_o a^\dagger_0 a_0 A^\dagger\cdot A  + \gamma_2 a^\dagger_0 a^\dagger_0 B\cdot B 
    + \gamma_2^*a_0 a_0 B^\dagger\cdot B^\dagger
    \label{eq:10}
\end{align}
where the $A$ and $B$ operators are collective bath operators defined by inspection of Eq.~\ref{eq:hfinal}. The term involving $A^\dagger\cdot A$ introduces
an energy fluctuation/dissipation simply due to scattering of the $k\ne 0$ population from the $k=0$ population.
The other two terms give rise to fluctuations/dissipation due to exciton pair creation/annihilation.
The constants $2 V_o$ and $\gamma_2$ can be determined by inspection of Eq.~\ref{eq:hfinal}.

To proceed, we shall drop the pair creation/annihilation 
terms and focus solely on the term involving 
$A^\dagger\cdot A$, we shall also treat these
as a collective variable and further assume that they undergo rapid thermalization via contact with a dissipative bath due to non-optical degrees of freedom. 
That is to say that we define a reservoir Hamiltonian
\begin{align}
    H_{res} &= \hbar\Omega (A^\dagger A + 1/2) + \sum_i g_i (b^\dagger_i A +  A^\dagger b_i)\nonumber \\
    &+ \sum_i\hbar\omega_i (b^\dagger_i b + 1/2)
\end{align}
Focusing solely upon the coupling to the reservoir, 
we define
\begin{align}
    \hat A(t) = A(t) e^{i\Omega t} \ \ \& \ \ \hat b_i(t) =  b_i(t)e^{i\omega_it}
\end{align}
so that the newly defined operators evolve only with the interaction. One obtained equations of motion of the form
\begin{align}
    i\hbar \partial_t \hat A &= \sum_i g_i \hat b_i(t)e^{i(\Omega-\omega_i)t} \\
    i\hbar \partial_t \hat b_i &= g_i \hat A(t)e^{-i(\Omega-\omega_i)t} \\
\end{align}
which can be integrated
\begin{align}
    \hat b_i(t)&= \hat b_i(t_o) - \frac{i}{\hbar}\sum_i g_i\int_{t_o}^{t-t_o}d t' \hat A(t') e^{i(\omega_i-\Omega) t'}
\end{align}
and inserted  to the equations of motion for $\hat{A}(t)$
\begin{align}
    \frac{d}{dt}\hat A(t) &= - \int_{0}^{t-t_0}d \tau \kappa(\tau) \hat A(t-\tau)  + \hat F(t)
\end{align}
where $\kappa(\tau)$ is given by 
\begin{align}
    \kappa(\tau) = \frac{1}{\hbar^2}\sum_i |g_i|^2 e^{i(\Omega-\omega_i)\tau}.
\end{align}
$\hat F(t)$ is a quantum operator acting on the reservoir variables:
\begin{align}
    \hat F(t) = -\frac{i}{\hbar}\sum_i g_i \hat b_i(t) e^{i(\Omega-\omega_i)t}
\end{align}
Integrating $\kappa(t)$ over all time 
\begin{align}
    \int_0^\infty \kappa(\tau)d\tau &= \frac{1}{\hbar^2}\lim_{\eta\to 0^+}\sum_i |g_i|^2 \int_0^\infty e^{i(\Omega-\omega_i + i\eta)\tau} d\tau\\
    &=\frac{1}{\hbar^2}\sum_i |g_i|^2 \left(\pi \delta(\Omega-\omega_i) + i {\cal P}\frac{1}{\Omega-\omega_i}\right) \\
    &= \frac{\gamma}{2} + i \Delta
\end{align}
where $\gamma$ is the spontaneous emission rate  and $\Delta$ is the energy shift.
Thus, the equation of motion for the collective $\hat A$ variables read
\begin{align}
    \partial_t \hat A = -\left(\frac{\gamma}{2} + i\Delta\right) \hat A + \hat F(t)
\end{align}
The $\hat F(t)$ remains a quantum operator that depends upon the reservoir variables 
and it is straightforward to write correlation functions assuming the
collective $\hat A$ operators are connected to thermal reservoir. 
\begin{align}
    \langle \hat F(t')\hat F(t) \rangle & =  \langle \hat F^\dagger(t')\hat F^\dagger(t) \rangle = 0\\ 
    \langle \hat F^\dagger(t')\hat F(t) \rangle & = \sum_i \frac{|g_i|^2}{\hbar^2}\langle n_i\rangle e^{i(\Omega - \omega_i)(t-t')}\\
    \langle \hat F(t')\hat F^\dagger(t) \rangle & =  \sum_i \frac{|g_i|^2}{\hbar^2}(\langle n_i\rangle +1)e^{i(\Omega - \omega_i)(t-t')}.
\end{align}

We now can take the background density as stochastic variable and re-cast Eq.~\ref{eq:10} as
\begin{align}
     H_0(t) =  \hbar\omega_0 a^\dagger_0a_0 +  \frac{V_0}{2} a^\dagger_0 a^\dagger_0a_0a_0 
    +2 V_o a^\dagger_0 a_0 N(t).
    \label{eq:27}
\end{align}
where by $\hbar=1$.
Converting to the interaction representation, the exciton operators evolves as
\begin{align}
    \hat{a}_0(t)=\exp{\left(-i\omega_0t-iV_0t\hat{n}_0 - i2 V_o \int_0^t N(\tau) d\tau \right)} \hat{a}_0 \equiv \hat U(t) \hat a_0,
\end{align}
where $\hat{n}_0=\hat{a}_0^\dagger\hat{a}_0$ is the number operator of $k=0$ excitons.

In order to integrate this we need to specify the initial conditions for the bath. Generally, one takes
it as being in a thermal state.  However, in the case we consider here, the background
excitations are generated by a laser pulse which creates a non-equilibrium
non-stationary ``bath'', characterized by an initial distribution related to the
power spectrum of the excitation pulse.
At this point we we shall take the background to be an {\em incoherent} population
 characterized by an initial mean $N_0$ and variance $\sigma_{N_0}$ 
 that evolved according to the stochastic differential equation
\begin{align}
    dN(t) = -\gamma N(t)dt + \sigma dW(t).
\end{align}
the variance $\sigma$ represents the equilibrium fluctuations (white noise) in the background population and  $dW(t)$ represents a Wiener process. 
The stochastic model is also called Ornstein-Uhlenbeck
model or mean-reverting model that describes a noisy relaxation process.\cite{OU1930,Fox1987,VonWeizsacker1990,Steele2001} 
The solution of the stochastic differential equation is 
\begin{equation}
     N(t)=N(0) e^{-\gamma t} + \sigma \int_0^t e^{-\gamma (t-s)} d W_s,
    \label{eq:nt}
\end{equation}
and the expectation value of the second term on the right hand side
is zero due to the property of the Brownian motion $W_s$. Setting $N_0=\langle N(0)\rangle$, the background population relaxes exponentially, 
\begin{align}
    \langle N(t)\rangle = e^{-\gamma t}N_0.
\end{align}
Formally, the spectral 
density of the bath can be obtained by Laplace 
transform of the
kernel in Eq.~\ref{eq:nt}, which implies that one 
can introduce more complex and nuanced stochastic
processes directly into the our model. 
Further, more complex kernels  do not present 
any limitation to our approach.  
The reason for this follows from the 
It{\^ o} isometry which states that for an arbitrary
adaptive 
process $f(t)$
\begin{equation}
    \nonumber \mathbb{E}\left[\left(\int_0^t f(s) dW_s\right)^2\right] = \mathbb{E}\left[\left(\int_0^t f(s)^2 ds \right)\right].
\end{equation}
where $\mathbb{E}[]$ denotes the expected value (i.e. average over noise).
Consequently, for an arbitrary stochastic process $N(t)$
specified by a stochastic differential equation of the form
\begin{align}
    dN = g(N,t) dt + f(N,t)dW(t)
\end{align}
we can use the It{\^ o} calculus to compute noise-averaged
expectation values.\cite{ito1944,biane2020,VonWeizsacker1990,Steele2001}  

\begin{widetext}

From this,  
the covariance of $N(s)$ and $N(t)$ can be computed as  
\begin{align}
    \nonumber \mathrm{Cov}(N_s,N_t) =&
    \mathbb{E}\left[(N_s-\mathbb{E}(N_s))(N_t-\mathbb{E}(N_t))\right] = \mathbb{E}(N_s N_t) - N_0^2 e^{-\gamma(t+s)}\\
    \nonumber =& \mathbb{E}\left[\sigma^2 \int_0^s e^{-\gamma(s-u)}dW_u \int_0^t e^{-\gamma(t-v)}dW_v \right] 
    + \mathbb{E} \left[ \left(N(0)-N_0\right)^2 e^{-\gamma(s+t)}\right] \\
    \nonumber & + \mathbb{E}\left[\sigma\left(N(0)-N_0\right)\left(e^{-\gamma s}\int_0^t e^{-\gamma(t-v)}dW_v + e^{-\gamma t}\int_0^s e^{-\gamma(s-u)}dW_u\right)\right].
    \label{eqn:covariance}
\end{align}
The first term on the right hand side can be simplified by the quadratic variation of It\^o calculus\cite{ito1944,biane2020,VonWeizsacker1990}
\begin{align}
    \nonumber & \sigma^2 e^{-\gamma(s+t)} \mathbb{E}\left[\left(\int_0^{\min(s,t)} e^{\gamma u} dW_u \right)^2 + \int_0^{\max(s,t)} e^{\gamma v} dW_v\right] \\
    &= \frac{\sigma^2}{2\gamma} e^{-\gamma(s+t)} \left(e^{2\gamma\min(s,t)}-1\right) = \frac{\sigma^2}{2\gamma} \left(e^{-\gamma|t-s|}-e^{-\gamma(t+s)}\right).
\end{align}
The second term reads $\sigma_{N_o}^2 e^{-\gamma (s+t)}$ with $\sigma_{N_o}^2$ being the variance of 
the background population at time $t=0$. 
The third term vanishes because the initially prepared background population $N(0)$ and the Wiener process $dW_t$ are statistically independent.
\subsection{Optical responses}
We can now use these results to derive the response functions for optical excitation.\cite{Mukamel:1995}
\begin{align}
     S^{(1)}(t) =& \frac{i}{\hbar} \langle [\hat \mu(t),\hat\mu(0)] \rho(-\infty)\rangle \\
     =& \frac{\mu^2}{\hbar} \left( \langle [\hat a^\dagger(t),\hat a(0)] \rho(-\infty)\rangle - c.c\right) \\
     =& -\frac{2\mu^2}{\hbar} \Im \left\langle \left( \exp\left[i\omega_0 t+iV_0 t \hat n_0 +i2 V_o\int_0^t  N(\tau) d\tau\right]\left[\left(e^{-iV_0t}-1\right)\hat n_0 -1\right]\rho(-\infty)\right) \right\rangle \\
     \nonumber &\approx -\frac{2\mu^2}{\hbar} \Im\left\{\left[\left(e^{-iV_0t}-1\right) n_0 -1\right] \exp\left[i(\omega_0 + V_0 n_0) t\right] \exp\left[i\frac{2 V_o}{\gamma}N_0\left(1-e^{-\gamma t}\right)\right] \right.\\
     &\left.\times \exp\left[-\frac{4 V_o^2\sigma^2}{4\gamma^3}\left(2\gamma t +4 e^{-\gamma t} -e^{-2\gamma t}-3\right) - \frac{4 V_o^2 \sigma_{N_o}^2}{2\gamma^2}\left(1-e^{-\gamma t}\right)^2 \right]\right\}.
     \label{eqn:S1}
\end{align}
in which the stochastic factor with $N(\tau)=N(0)+dN(\tau)$ is subjected to the second-order cumulant expansion,
\begin{align}
    \left\langle \exp{\left[-i2 V_o \int_0^t N(\tau)d\tau\right]}\right\rangle = \exp{\left[i2 V_o g_1(t)\right]}\exp{\left[-2V_o^2 g_2(t)\right]} \cdots.
\end{align}
The cumulants read
\begin{align}
    g_1(t) &= \int_0^t \langle N(\tau) \rangle d\tau =\frac{N_0}{\gamma}\left(1-e^{-\gamma t}\right), 
    \label{eqn:g1}\\
    \nonumber
    g_2(t) &= \int_0^t \int_0^{t} \mathrm{Cov}\left[N(\tau),N(\tau')\right] d\tau' d\tau \\
    &= \frac{\sigma^2}{2\gamma^3}\left(2\gamma t + 4e^{-\gamma t} - e^{-2\gamma t} - 3\right) + \frac{\sigma_{N_o}^2}{\gamma^2}\left(1-e^{-\gamma t}\right)^2,
    \label{eqn:g2}
\end{align}
where the initial background population averages $\langle N(0)\rangle=N_0$ and has the variance of $\sigma_{N_o}^2$.
The expressions for $g_1(t)$ and $g_2(t)$ constitute  
the central results of this paper. 
For completion, we note that the double time integral in this last expression can also 
be obtain for when the two time limits are not equal. 
\begin{align}
    \nonumber \int_0^t \int_0^{t'} \mathrm{Cov}& \left[N(\tau), N(\tau')\right] d\tau' d\tau = \\
    &\frac{\sigma^2}{2\gamma^3}\left[2\gamma~ \mathrm{min}(t,t') + 2e^{-\gamma t} + 2e^{-\gamma t'} - e^{-\gamma|t'-t|} - e^{-\gamma(t'+t)} -2\right] \\
    \nonumber &+ \frac{\sigma_{N_o}^2}{\gamma^2}\left[e^{-\gamma(t+t')}-e^{-\gamma t}-e^{-\gamma t'}+1\right].
\end{align}
This term will arise in the analysis of the non-linear responses as we discuss in Ref.~\citenum{paper1}.
\end{widetext}
\subsection{Comparison to Anderson-Kubo theory}

The expressions we have above are exact insofar as the assumptions of our 
stochastic model is concerned.  We now compare our approach to the more familiar
Anderson-Kubo model in order to point out some key similarities and crucial differences. 
First, let us assume that the background population follows a stochastic process
given by 
\begin{align}
    N(t) = N_s + \delta N(t)
\end{align}
where $N_s$ is the {\em stationary} background population and $\delta N(t)$ 
corresponds to fluctuations about that stationary state. 
In principle, $N_s$ can be set to zero, but we shall carry it through as non-zero until 
the end. 
Following AK, we can write the variance as 
\begin{align}
    \langle \delta N(t) \delta N(0)\rangle = \sigma_K^2 e^{-\gamma t}.
    \label{eq:AKmodel}
\end{align}
Note, that we shall use $\sigma_K$ to discriminate between the variance in the 
population in the AK model versus the variance $\sigma^2$ in Eq.~\ref{eq:nt}.
The distinction is crucial since $\sigma_K^2$ is unitless 
while the $\sigma^2$ in the Ornstein-Uhlenbeck stochastic differential 
equation (Eq.~\ref{eq:nt}) carries units of $[t^{-1}]$.
Within the context of AK, 
\begin{widetext}
\begin{align}
    \left\langle \exp\left[-2\frac{i}{\hbar}V_o \int_0^t N(\tau)d\tau\right]\right\rangle &= \exp\left(\frac{i}{\hbar}2 V_o N_s t\right)\exp\left[-\frac{4 V_o^2 \sigma_{K}^2}{\hbar^2 \gamma^2}\left(e^{-\gamma t}+\gamma t-1\right) \right] \\
    &= \exp\left(\frac{i}{\hbar}2 V_o N_s t\right)\exp\left[-\frac{4 V_o^2}{\hbar^2} g_2^{\rm K}(t) \right]. \label{eq:AKcumulant}
\end{align}
\end{widetext}
Here, the background population produces a frequency shift proportional to the mean-field
interaction strength, $2 V_o N_s$, as well as the usual AK lineshape function
\begin{align}
    g_{2}^{\rm K}(t) = \frac{\sigma_{K}^2}{\gamma^2}(e^{-\gamma t} + \gamma t -1).
\end{align}

\begin{figure}
    \centering
    \includegraphics[width=0.95\columnwidth]{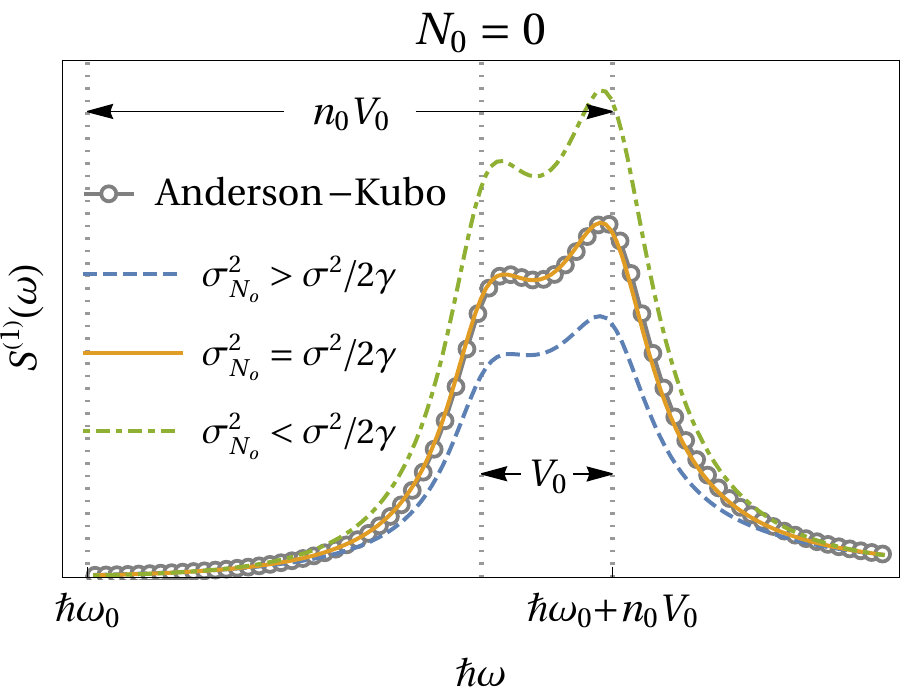}
    \caption{The linear response function comparison between the non-stationary and the Anderson-Kubo (AK) 
    model in the case of zero initial background population $N_0$ at different distributions $\sigma_N^2=0.25$, 0.125, and $0.04~{\rm fs}^{-1}$. Other parameters are $V_o=10~{\rm meV}$, $\gamma=0.01~{\rm fs}^{-1}$, $\sigma^2=0.0025~{\rm fs}^{-1}$. }
    
    \label{fig:fig1}
\end{figure}

Fig.~\ref{fig:fig1} we show the  linear response 
for the case where no initial background excitations
are produced at time $t=0$ ($N_0=0$), but background
fluctuations are present.  We compare the present results 
against the Anderson-Kubo model, which we will discuss in the
next section. For reference, we set the 
energy origin to the bare exciton energy $\hbar\omega_o = 2.35$eV
to correspond to the exciton energies of the \ce{(PEA)2PbI4} hybrid perovskite
system.   Our goal here is not so much to reproduce the experimental spectra, 
but to understand how the line-shape changes in 
a physically relevant parametric regime.
With this in mind, we 
set the exciton/exciton interaction $V_o = 10 {\rm meV}$, 
relaxation rate $\gamma = 0.01 {\rm fs}^{-1}$, 
and noise variance $\sigma^2 = 0.0025 {\rm fs}^{-1}$ unless indicated
otherwise.  
First, the center peak is shifted relative to the
bare exciton towards higher energies due to 
exciton/exciton self-interactions. 
The two peaks arise from the
$
(e^{-iV_0t}-1) n_0
 $
 contribution in Eq.~\ref{eqn:S1} and scale with increasing $n_0$ population. 
 The asymmetry arises from the $g_2(t)$ line shape function.  

The expression for $g_2(t)$ in Eq.~\ref{eqn:g2} can be rearranged to read
\begin{widetext}
\begin{align}
    g_2(t) = \left(\frac{\sigma_{N_o}^2}{\gamma
   ^2}-\frac{3 \sigma ^2}{2 \gamma ^3}\right)
   +\frac{2}{\gamma^3}e^{-\gamma t} \left(\sigma ^2-\gamma \sigma_{N_o}^2\right)
   -\frac{e^{-2 \gamma t}}{2\gamma^3}
   \left({\sigma ^2}-2\gamma\sigma_{N_o}^2\right)   
   +\frac{\sigma ^2 }{\gamma ^3} t 
    \end{align}
\end{widetext}
which allows us to introduce $\kappa = \sigma^2-2\gamma\sigma_{N_0}^2$ as a 
important parameter in determining the overall spectral lineshape
When $\kappa \ne 0$, the dynamics are determined by magnitude of the 
Brownian noise that characterizes the steady-state of the background. 
This is equivalent to stating that the initial excitation is narrow ($\kappa > 0$)
or broad ($\kappa < 0$)
compared to the background fluctuations.  
When $\kappa = 0$, we obtain an important limiting case of our model. 
In this specific case, our model is {\em exactly} equivalent to the 
Anderson-Kubo model {\em viz.}
\begin{align}
    g_2(t) \to \frac{\sigma^2}{\gamma^3}\left(e^{-\gamma t} + \gamma t -1\right).
\end{align}
giving $\sigma^2_K \equiv \sigma^2/\gamma$.
This limit only holds when the initial excitation pulse produces $N_0 = 0$ 
and that the fluctuations are about the steady-state.

\subsection{Non-stationary spectra.}

\begin{figure*}
    \centering
    \subfigure{\includegraphics[width=0.45\textwidth]{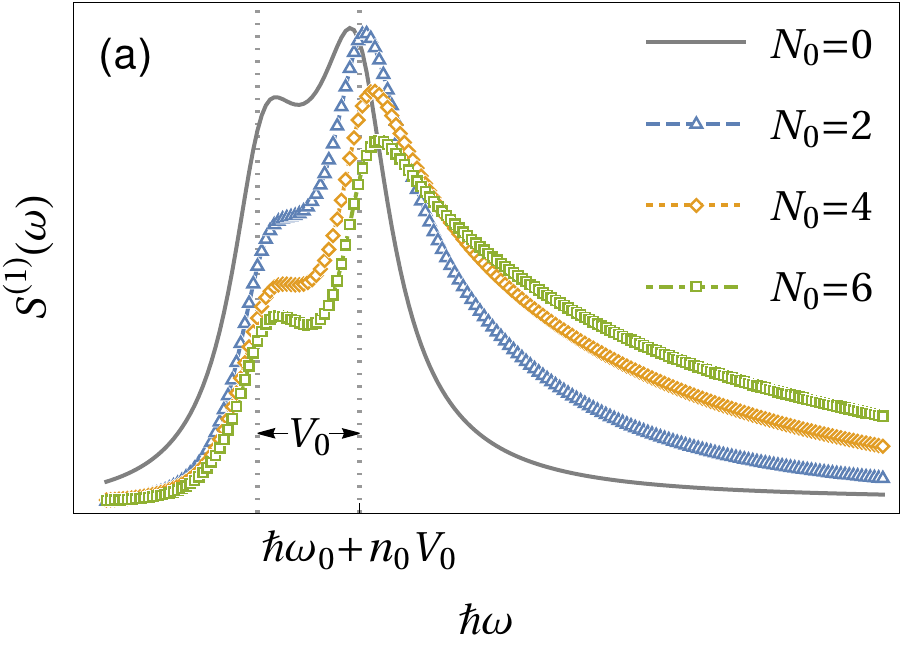}}
    \subfigure{\includegraphics[width=0.45\textwidth]{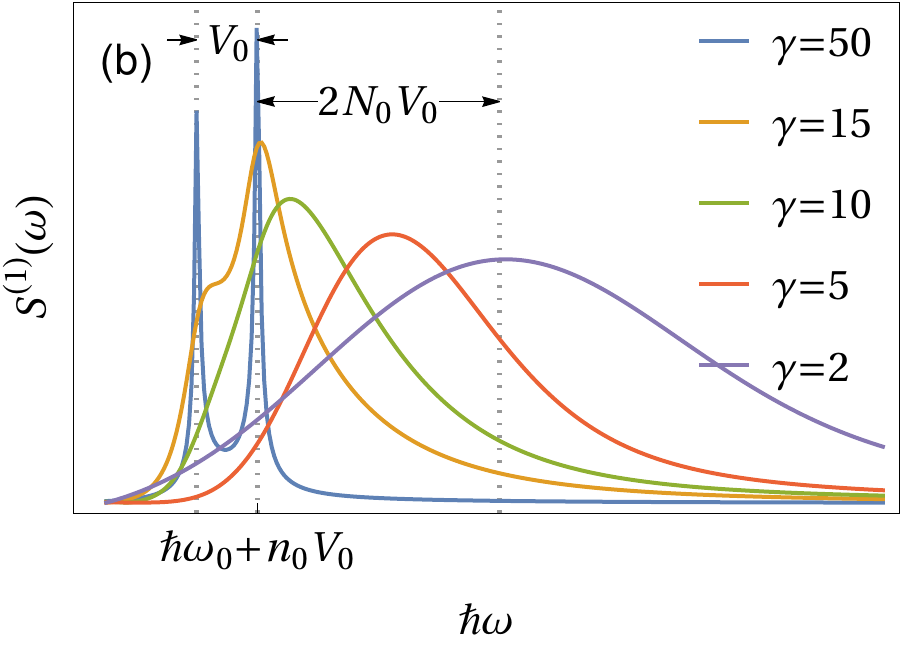}}\\
    \caption{The linear response function with (a) increasing background population density $N_0$, and (b) different relaxation rate $\gamma$, from the homogeneous limit of $\gamma=50$ meV to the inhomogeneous limit of $\gamma=2$ meV.}
    \label{fig:fig2a-d}
\end{figure*}

When the excitation pulse produces a non-stationary 
background population ($N_0> 0$), it is no longer consistent 
to compare against the Anderson-Kubo model, which is valid solely 
for a stationary background. 
Fig. \ref{fig:fig2a-d}(a) displays the effect of a non-stationary 
background on the linear absorption spectrum of a system. 
The notable feature is the tail that extends to higher absorption 
energies.  The character of this tail depends most strongly upon 
the initial choice of $N_0$ and is attributable to the $g_1(t)$ term
in our response function which is the time-integral over the 
evolving background population.  This term, as it
appears in Eq.~\ref{eqn:S1} produces an {\em evolving} frequency shift
reflecting the dynamical relaxation of the background.   In the 
$S^{(1)}$ response, this produces a tail extending out to the blue. 
Our model exhibits all the correct features observed
in the absorption spectroscopy of typical 2D semiconductor systems 
and transition metal dichalcogenides.~\cite{Katsch2020}
\begin{enumerate}
    \item {\em Blocking}: Increasing the initial background suppresses the peak 
    absorption intensity. 
     \item {\em Energy shift}: The peak position shifts to the blue with increasing
     background population due to increased Coulombic interactions. 
     \item {\em Broadening:} The spectrum acquires a long tail extending to the 
     blue due to the dynamical evolution of the background.  
     This feature also appears in the 2D coherent spectroscopy as an asymmetry
     along the absorption axis and as phase scrambling in the rephasing and non-rephasing 
     signals.~\cite{paper1}
     \item {\em Biexciton:} The peak is split by $V_0$ corresponding to the 
     biexciton interaction. 
\end{enumerate}
In Ref.~\citenum{paper1} where we compute the non-linear coherent responses, 
we find that this evolution produces both asymmetry
as well as phase scrambling  in the 2D spectroscopy 
of \ce{(PEA)2PbI4}.~\cite{thouin2019enhanced}

In Fig.~\ref{fig:fig2a-d} we compare the effect of decreasing the 
relaxation rate $\gamma$ for fixed values of $N_0=4$ which 
carries the system from the homogeneous limit ($\gamma = 50$meV) 
in which the background relaxation is very fast to the 
fully broadened inhomogeneous limit.  Under this 
extreme, the exciton and bi-exciton spitting is clearly resolved
and the lineshapes are Lorenzian about each peak. 
to the inhomogenous limit.  Decreasing the relaxation rate $\gamma$
produces a systematic shift towards the blue due to the mean-field
interaction between the exciton and the background.  This shift
saturates when the peak is fully shifted by $2V_0N_0$ and acquires 
a Gaussian form reflecting mean $N_0$ and variance $\sigma_{N_o}$ of 
the initial background.  Physically this corresponds to the 
the excitation pulse as projected onto the 
density of states of the system.  

At this point it is worth comparing our results and model to 
the work recently presented by Katsch, et al. in Ref.\citenum{Katsch2020}.
In this the authors use a many-body/Heisenberg equations of motion 
approach to describe inter-valley electron/hole interactions 
in 2d semiconductors systems and use this approach to compute  the
response to an time-dependent electric field explicitly incorporated
into the equations of motions.  In principle, this is an exact calculation
and provides a highly useful benchmark for the present theoretical analysis.
Our analytical model produces all the important features 
found in the more detailed computational approach and offers 
additional insight into the underlying physics leading to these effects. 
Moreover, our analytical model provides an efficient avenue for 
computing higher-order coherent spectroscopies, 
which we present in Ref.~\citenum{paper1}.

\section{Discussion}

We present here what appeared to be a straightforward extension of the Anderson-Kubo
model in the sense that the external environment serves as a non-stationary source
of noise which modulates the energy gap of a given transition. 
We find that model produces non-linear spectral shifts and asymmetries 
that depend vary systematically with the initial background population
and are consistent with the absorption characteristics of 
a wide range of semiconducting systems.  
In the most general 
sense, our model does not hinge upon a specific model for the 
the environmental noise, we only require that it follow from stochastic differential 
equation that can be integrated using It\^o's Lemma.
In principle, one can implement more system-specific noise-sources, 
as well as correlated sources, directly into our approach. 

In the accompanying paper (Ref.~\citenum{paper1}) we explore the implications of the
model for higher-order non-linear coherent optical responses and apply the 
approach to study the excitation-induced dephasing (EID) observed in a
hybrid perovskite semiconductor.

\begin{acknowledgments}
The work at the University of Houston was funded in
part by the  National Science Foundation (
CHE-1664971,   
DMR-1903785    
) and the Robert A. Welch Foundation (E-1337). 
The work at Georgia Tech was funded by the National Science Foundation (DMR-1904293). CS acknowledges support from the School of Chemistry and Biochemistry and the College of Science at Georgia Tech. 
 
\end{acknowledgments}

\section*{Data Availability}

The data that support the findings of this study are available from the corresponding author upon reasonable request.


%

\end{document}